\documentclass{ifacconf}

\usepackage{graphicx, amsmath, amsfonts, amssymb, mathtools, enumerate, color}      
\usepackage{natbib}        

\newtheorem{setting}{Setting}
\newtheorem{definition}{Definition}
\newtheorem{theorem}{Theorem}

\newtheorem{propn}{Proposition}
\newtheorem{assumption}{Assumption}

\begin{document}
\begin{frontmatter}

%

\title{Influencing Opinion Dynamics in Networks with Limited Interaction} 

\thanks[footnoteinfo]{This work was supported in part by an Indo-French grant on “Machine Learning for Network Analytics". The work of Neeraja Sahasrabudhe was also supported in part by the DST-INSPIRE Faculty Fellowship from the Govt. of India. The work of Sharayu Moharir was supported in part by a seed grant from IIT Bombay.}

\author[First]{Anmol Gupta} 
\author[First]{Sharayu Moharir} 
\author[Third]{Neeraja Sahasrabudhe}

\address[First]{Indian Institute of Technology Bombay}
\address[Third]{Indian Institute of Science Education and Research Mohali}

\begin{abstract}                
The focus of this work is on designing influencing strategies to shape the collective opinion of a network of individuals. We consider a variant of the voter model where opinions evolve in one of two ways. In the absence of external influence, opinions evolve via interactions between individuals in the network, while, in the presence of external influence, opinions shift in the direction preferred by the influencer. We focus on a finite time-horizon and an influencing strategy is characterized by when it exerts influence in this time-horizon given its budget constraints. Prior work on this opinion dynamics model assumes that individuals take into account the opinion of all individuals in the network. We generalize this and consider the setting where the opinion evolution of an individual depends on a limited collection of opinions from the network. We characterize the nature of optimal influencing strategies as a function of the way in which this collection of opinions is formed.


\end{abstract}


\end{frontmatter}

\section{Introduction}
The field of opinion dynamics has attracted a lot of attention across domains including physics, mathematics, and epidemiology. A popular and widely studied binary opinion dynamics model is the voter model (\cite{clifford1973model, holley1975ergodic}).
Many variants of the voter model have been proposed and studied, for instance, the linear threshold model proposed in \cite{kempe2003spread}, the majority model proposed in \cite{gartner2017majority}, the conformist model and the strong-willed model studied in \cite{MTNS}, and the rebel model studied in \cite{TCNS}. 

Motivated by applications like advertising and political campaigning, designing influencing strategies to shape opinions in networks is another area that has seen a lot of work, e.g., \cite{kandhway2014run, kotnis2017incentivized, kempe2005influential, eshghi2017spread, MTNS}. Refer to \cite{eshghi2017spread} for a comprehensive survey of works in this field. A related body of research focuses on designing strategies to prevent/slow down the spread of a disease in a network includes works like \cite{asano2008optimal, ledzewicz2011optimal, lashari2012optimal}.

In this work, we consider two modes of opinion evolution, namely, organic evolution like in the voter model due to interactions with peers, and evolution due to external influence which attempts to move the network towards the favored opinion. Closest to our work, \cite{MTNS} and \cite{TCNS} also consider this hybrid opinion dynamics setting. Time is divided into slots and an individual is chosen uniformly at random in each time-slot. Only the chosen individual can change her opinion. In the absence of external influence, the opinion of the individual chosen in a time-slot changes with probability proportional to the fraction of individuals in the network with the opposing opinion at that time. In time-slots during which external influence is exerted, the opinion of the chosen individual is more likely to move in the direction preferred by the external influence than in time-slots without external influence. Given an influencing budget and a fixed time-horizon, the goal is to determine when to influence within the time-horizon to maximize the fraction of individuals with the preferred opinion at the end of the time-horizon. The key takeaway from \cite{MTNS} and \cite{TCNS} is that influencing at the beginning of the time horizon is more effective than influencing at the end only when individuals with a favorable opinion are less likely to change their opinion than individuals with an unfavorable opinion. In all other cases, the conventional wisdom of advertising towards the end of the time-horizon is indeed an optimal strategy. 

We generalize the setting studied in \cite{MTNS} and \cite{TCNS} by considering the case where, in each time-slot without external influence, the chosen individual changes their opinion with probability proportional to the fraction of individuals with the opposite opinion in a collection of opinions sought by the chosen individual. Unlike the setting in \cite{MTNS} and \cite{TCNS}, this collection need not include the opinions of all the individuals in the network. 

Our goal is to explore if the nature of optimal influencing strategies obtained in \cite{MTNS} and \cite{TCNS} are robust to these modifications. 
We focus on two specific ways through which this collection is curated. 

In the first case, the chosen individual samples a random number of opinions, each chosen uniformly at random from the set of opinions in the network. In this case, we show that the optimal influencing strategies remain identical to the case studied in \cite{MTNS} and \cite{TCNS}. This result holds independent of the probability distribution of the number of opinions the chosen individual seeks.

In the second case,  we use a fixed graph to model links between individuals. Each chosen individual only takes into account the opinion of her neighbors in this graph. Here, the nature of the optimal influencing strategies depends on the nature of the graph. For popular network models like $d$--regular graphs, Barab\'{a}si Albert graphs and Erd\"{o}s R\'{e}nyi graphs, the optimal influencing strategies are identical to those in the case of the complete graph. An example where the nature of optimal strategies is not the same as that in for the complete graph is the hub and spoke graph where a single node, called the hub has links to all other nodes in the network, while the remaining nodes are only connected to the hub.   

\section{Setting} \label{sec:settings}
\subsection{Notation and System Evolution}
We consider a finite population of $M$ individuals where every individual has a binary opinion (Yes or No) about a specific topic. We study the evolution of the cumulative opinion of the population in the presence of an external influencing agency with a limited budget. This agency attempts to maximize the number of individuals with opinion Yes at the end of a fixed finite time horizon.

We consider a population of size $M$ and index the individuals as $\{1,...,M\}$. We divide time into slots of equal duration. Random variables \{$I_i(t)$\}$_{1 \leq i \leq M, t \geq 0}$ taking values in $\{0,1\}$, denote the opinion of the $i^{\text{th}}$ individual at time $t$, such that
$$
  I_i(t) = 
\begin{cases}
1, \mbox{ if the opinion of Individual $i$ at time $t$ is Yes} \\
0, \mbox{ if the opinion of Individual $i$ at time $t$ is No}.
\end{cases}
$$
At time $t$, let $\beta(t)$ and $\delta(t) = 1-\beta(t)$ be the fraction of people with opinion Yes and No respectively. 

In time-slot $t$, an individual is selected uniformly at random and she updates her opinion as follows: the selected individual, say Individual $i$, collects opinions from a group of individuals in the population. The opinion of the Individual $i$ is then updated based on the collective opinion in the group. Let $\widetilde{\beta_i}(t)$ and $\widetilde{\delta_i}(t)$ denote the fraction of Yes and No opinions in the collection. We consider two ways in which this collection of opinions is formed. 


\begin{setting}[Random local interactions] 
\label{setting:random}
The selected individual collects $K$ opinions, each chosen uniformly at random from the set of $M$ opinions. Note that there can be repetitions in the collected opinions. We model $K$ as a random variable. 
\end{setting}
\begin{setting}[Graph-based local interactions]
\label{setting:fixed}
In this case, we consider a fixed graph $G=(V, E)$ such that $|V|=M$. We think of individuals as the nodes of the graph. Define $N_i = \{ j : (i,j) \in E \}$ as the neighbourhood of $i^{\text{th}}$ individual. Each individual $i$, if selected, collects opinions from her neighbors $j \in N_i$. 
\end{setting}

The opinion of Individual $i$ chosen in time-slot $t$ evolves according to a Markov process with the following transition probabilities.
\begin{align*}
    P(I_{i}(t+1) = 0 | I_{i}(t) = 1) &= p_t\\
    P(I_{i}(t+1) = 1 | I_{i}(t) = 1) &= 1 - p_t\\
    P(I_{i}(t+1) = 1 | I_{i}(t) = 0) &= q_t\\
    P(I_{i}(t+1) = 0 | I_{i}(t) = 0) &= 1 - q_t.
\end{align*}
The values of $p_t$ and $q_t$ depend on whether the individual is being externally influenced in time-slot $t$ or not. More specifically,
\begin{enumerate}
    \item[--] If the chosen individual is being externally influenced in time-slot $t$, $p_t = \tilde{p}$ and $q_t = \tilde{q}$ for some fixed $\tilde{p}, \tilde{q} \geq 0$. 
    \item[--] Else, the probability of an individual changing her opinion is proportional to the fraction of contrary opinions in the collection, i.e., $p_t = p\widetilde{\delta_i}(t)$ and $q_t = q \widetilde{\beta_i}(t)$ for some fixed $p, q \geq 0$.
\end{enumerate}

\begin{assumption}[Effective Influence]
	\label{ass:effective}
	If the chosen individual has opinion No(/Yes), then the probability of the individual switching their opinion to Yes(/No) under external influence is more(/less) compared to its value in the absence of external influence, irrespective of the state of opinions in the network. 
	Formally, $\tilde{p} < p$ and $\tilde{q} > q$. 
\end{assumption}

\subsection{Goal}
We focus on a finite time-horizon of $T$ consecutive time-slots, indexed $1, 2, \cdots, T$. Due to budget constraints, the external influencing agency can influence in at most $bT$, $b \in [0,1]$ out of the $T$ time-slots. The goal is to determine when to exert external influence in order to maximize the fractions of individuals with opinion Yes at the end of time-slot $T$. 

\begin{definition}[Optimal Strategy] \label{OS}
	An influence strategy is characterized by the indices of the time-slots in which external influence is exerted. We call an influence strategy optimal if it results in at least as many expected number of individuals with opinion Yes at the end of time $T$ as under any other strategy.
\end{definition}

\section{Preliminaries} \label{sec:prelim}
In this section, we introduce the mathematical framework used in the rest of this paper. 



Suppose Individual $i$ is chosen at time $t$. She updates her opinion as follows:
\begin{equation}
I_i(t+1) = I_i(t) + \xi_i(t+1),
\end{equation}
where, $\xi_i(t + 1)$ is a random variable taking values in $\{-1, 0, 1\}$ and denotes the change in the opinion of Individual $i$ in time-slot $t$.
We are interested in the fraction of individuals with opinion Yes in the population given by $\beta(t) = \dfrac{1}{M} \sum\limits_{i=1}^M I_i(t)$, which evolves as follows:
\begin{equation}\label{beta} \beta (t+1) = \beta (t) + \frac{1}{M} \xi (t+1), \end{equation}
where $\xi (t+1) \in \{-1, 0, 1 \}$ denotes the change in net opinion at time $t$. Thus, conditioned on the event that Individual $i$ is selected at time $t$, $\xi (t+1) = \xi_i(t+1)$. Let $\mathcal{F}_t$ denote the $\sigma$-field generated by the random variables $\{\xi(1), \xi(2), ... , \xi(t) \}$. The evolution of the opinion is governed by the random process $\xi(t)$. Since the evolution depends on whether the selected individual has opinion Yes or No at time $t$, we have:
\begin{align*}
\begin{split}
    P(\xi(t+1) = x | F_t) = \mspace{250mu} \\
    \begin{cases}
        \beta (t) p_t &\text{for } x = -1\\
        (1 - \beta (t)) q_t &\text{for } x = 1\\
        1 - q_t - \beta (t)(p_t - q_t) &\text{for } x = 0.
    \end{cases}
\end{split}
\end{align*}
We rewrite \eqref{beta} as:
\begin{eqnarray*}
    \beta(t+1) &=& \beta(t) + \frac{1}{M} E[\xi (t+1) | \mathcal{F}_t] \\
    &&+ \frac{1}{M}[\xi (t+1) - E[\xi (t+1) | \mathcal{F}_t]], 
\end{eqnarray*}
where, $\xi (t+1) - E[\xi (t+1) | \mathcal{F}_t]$ is a Martingale difference sequence. 

To analyze the evolution of random variable $\beta(t)$, we use the theory of constant step-size stochastic approximation. A constant step-size stochastic approximation scheme is given by the following iteration for $ x \in \mathbb{R}^d $ and $a>0$. Formally,
\begin{equation} \label{stoch}
    x_{n+1} = x_n + a[h(x_n) + M_{n+1}], n \geq 0,
\end{equation}
such that:
\begin{enumerate}[(i)]
    \item[--] $ h : \mathbb{R} \rightarrow \mathbb{R}^d $ is Lipschitz.
    \item[--] $ \{M_n\}_{n \geq 0}$ is a square-integrable Martingale difference sequence with respect to a suitable filtration.
    \item[--]  $\{ \| x_n \|^2 \}$ are uniformly integrable.
    \item[--] $\sup_n E[\| x_n \|^2]^{1/2} < \infty$.
\end{enumerate}
Then, from the theory of stochastic approximation in \cite{borkar2008stochastic} we know that the iterates of \eqref{stoch} track the following O.D.E. with high probability:
\begin{equation}
 \label{ODE}
    \dot{x}(t) = h(x(t)), t \geq 0.
\end{equation}
Specific bounds can be found in Lemma 1, Chapter 9 and Theorem 3, Chapter 9 in \cite{borkar2008stochastic}. 

Given this, for each setting we consider, we characterize \eqref{stoch} for $\beta(t)$ and solve the O.D.E. as given in \eqref{ODE}. Since the solutions of the recursion and that of the O.D.E. remain close with high probability, it suffices to study the optimal strategies for the corresponding O.D.E. system. This approach is widely used and accepted, including in \cite{MTNS} and \cite{TCNS}.


\section{Main Results and Discussion} \label{sec:Results}
In this section, we present our results. We consider the two settings discussed in Section \ref{sec:settings} separately. The proofs are presented in Section \ref{sec:Proofs}.
\subsection{Setting \ref{setting:random}: Random Local Interactions}
Our first result characterizes the optimal policy for Setting~\ref{setting:random}.
\begin{theorem}
Under Assumption \ref{ass:effective}, for Setting \ref{setting:random} discussed in Section \ref{sec:settings},
\begin{itemize}
	\item[--] if $p < q$, an optimal strategy is to influence in the first $bT$ time-slots of the finite time-horizon and the strategy of influencing in the last $bT$ time-slots is strictly sub-optimal. 
	\item[--] if $p > q$, an optimal strategy is to influence in the last $bT$ time-slots of the finite time-horizon and the strategy of influencing in the first $bT$ time-slots is strictly sub-optimal. 
	\item[--] if $p = q$, all policies which influence in any $bT$ out of the $T$ time-slots in the finite time-horizon perform equally well.
\end{itemize}
\end{theorem}
We note that this result holds independent of the distribution of $K$, values of $M$, $T$, $b$ and the initial conditions. This is identical to the result obtained in Theorem 3 in \cite{MTNS}. We thus conclude that the nature of the optimal strategy does not change if the chosen individual uses a random sampling of opinions in the network to shape her opinion instead of taking into account the opinions of all the individuals, as is done in \cite{MTNS}. 


\subsection{Setting \ref{setting:fixed}: Graph-based local interactions}
We now focus on the second setting described in Section \ref{sec:settings} where there is a graph connecting the $M$ individuals in the network and each individual only takes into account her neighbors' opinions. 

\subsubsection{(a) $d$--regular graphs:}
Our next result characterizes the optimal policy when the graph $G(V,E)$ is $d$--regular.

\begin{theorem}
	\label{thm:dRegular}
Under Assumption \ref{ass:effective}, for Setting \ref{setting:fixed} discussed in Section \ref{sec:settings} if the graph $G$ is $d$--regular, we have:
\begin{itemize}
	    \item[--] if $p < q$, an optimal strategy is to influence in the first $bT$ time-slots of the finite time-horizon and the strategy of influencing in the last $bT$ time-slots is strictly sub-optimal. 
    \item[--] if $p > q$, an optimal strategy is to influence in the last $bT$ time-slots of the finite time-horizon and the strategy of influencing in the first $bT$ time-slots is strictly sub-optimal. 
    \item[--] if $p = q$, all policies which influence in any $bT$ out of the $T$ time-slots in the finite time-horizon perform equally well.
\end{itemize}
\end{theorem}
We note that this result holds independent of the value of $d$, $M$, $T$, $b$ and the initial conditions. 
We thus conclude that the results obtained in \cite{MTNS} for the complete graph extend to $d$--regular graphs.

\subsubsection{(b) Barab\'{a}si Albert and Erd\"{o}s R\'{e}nyi graphs:}
\begin{figure}[t]
	\centering
	\includegraphics[scale = 0.78]{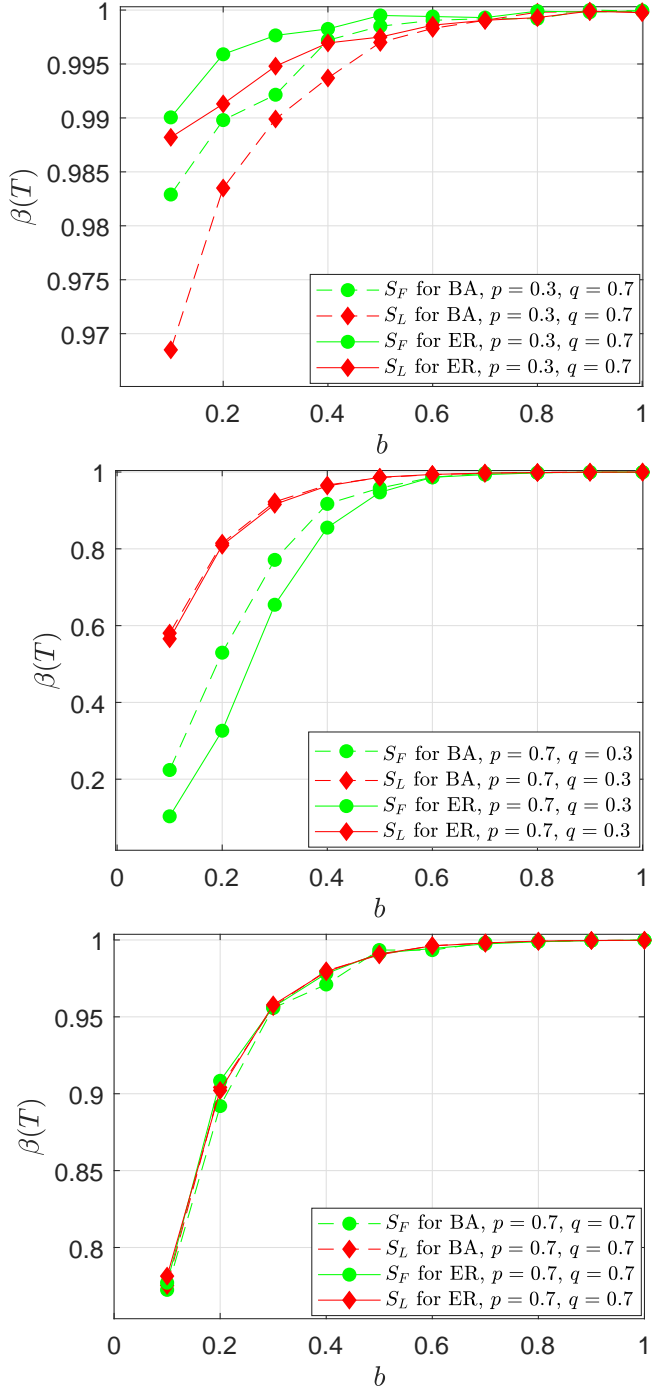}
	\caption{BA and ER Graphs: Comparison between the strategy to influence in the first and last $bT$ time-slots ($S_F$ and $S_L$ respectively)  where $M = 100, T = 500$, $\tilde{p} = 0$, and $ \tilde{q} = 0.75$. The performance follows the same trend as for $k$--regular graphs (Theorem \ref{thm:dRegular})}
	\label{fig:figure2}
\end{figure}

In Figure~\ref{fig:figure2}, via simulations, we show that the same trends are observed in Barab\'{a}si Albert and Erd\"{o}s R\'{e}nyi graphs. The models are among the most widely studied random graphs models. In addition, Barab\'{a}si Albert graphs have a power-law degree distribution and are therefore considered suitable to model many networks including genetic networks, the World Wide Web and social networks as discussed in \cite{barabasi1999emergence}. As is the case for complete and $d$--regular graphs, the strategy to influence in the beginning ($S_F$) outperforms the strategy to influence at the end ($S_L$)  if $p<q$ and the opposite is true of $p>q$. For $p=q$, the performance of the two policies is indistinguishable. 




%


Next, we present an example of a graph for which the trend of the performance of the policies discussed in Theorem \ref{thm:dRegular} is different from that for $d$--regular, Barab\'{a}si Albert, Erd\"{o}s R\'{e}nyi and complete graphs.

\subsubsection{(c) Hub and Spoke Graphs: }
A hub and spoke graph is a graph $G=(V, E)$ with $|V|=M$, such that one vertex called the hub, has degree $M-1$, while all the other vertices have degree $1$ and are only connected to the hub.
The extreme variation in the degrees of nodes in the graph introduces a high disparity between the influence that the two kinds of nodes have on the overall opinion dynamics.


\begin{figure}[t]
    \centering
    \includegraphics[scale = 0.78]{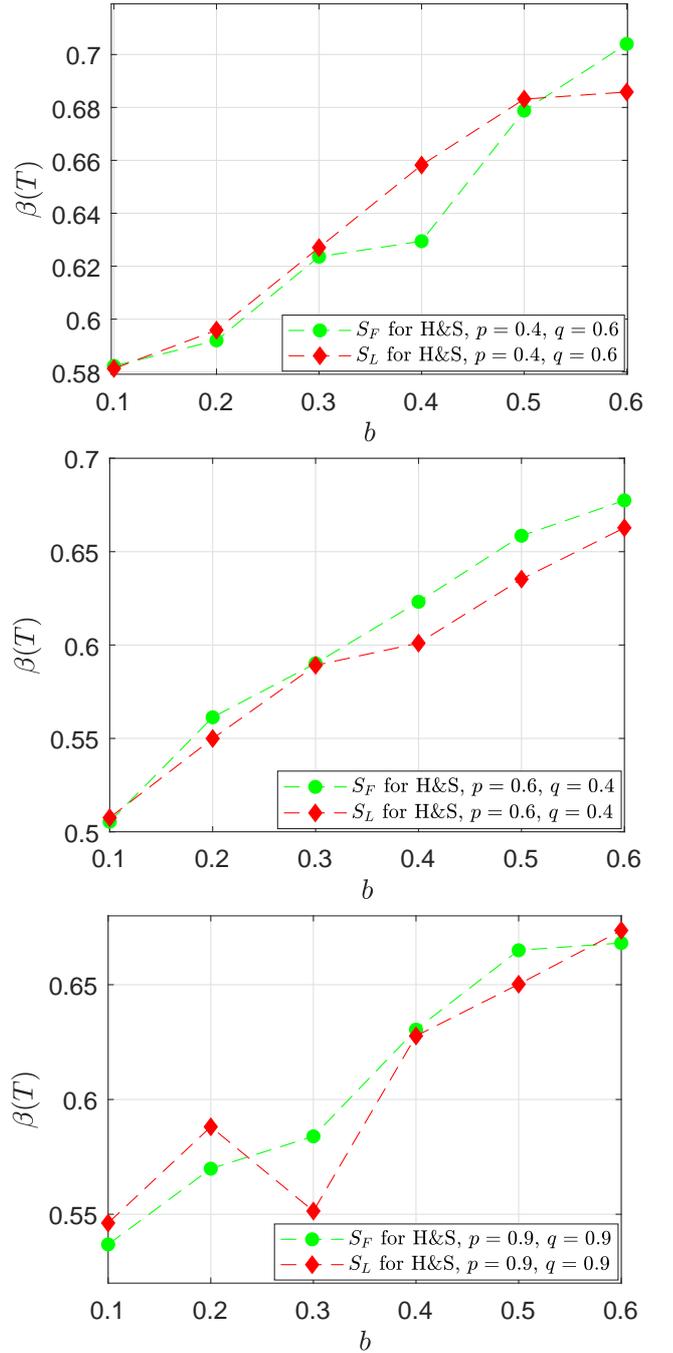}
    \caption{Hub and Spoke Graphs: Comparison between the strategy to influence in the first and last $bT$ time-slots ($S_F$ and $S_L$ respectively)  where $M = 500, T = 500$, $\tilde{p} = 0$, and $\tilde{q} = 0.75$. The performance does not follow the same trend as for $k$--regular graphs (Theorem \ref{thm:dRegular}) and BA and ER graphs (Figure \ref{fig:figure2})}
    \label{fig:figure5}
\end{figure}


In Figure \ref{fig:figure5}, we compare the strategy to influence in the beginning ($S_F$) and the strategy to influence at the end ($S_L$) for the hub and spoke graph. We conclude that unlike the graph models discussed above, there is no clear trend in the performance of these two policies for the hub and spoke graph.

An exact characterization of the optimal influence strategy for this case is analytically challenging and therefore remains an open problem. In our next proposition, we provide some insight into the performance of various influence strategies under some special cases.  

%

\begin{propn}
	Under Assumption \ref{ass:effective}, for Setting \ref{setting:fixed} discussed in Section \ref{sec:settings} if the graph $G$ is a hub and spoke graph,
	\begin{enumerate}
		\item[--] Consider the case when the hub is not selected in $t \in [1, T]$. If the hub has opinion Yes at $t=1$, all strategies perform equally well for all values of $p$ and $q$. If hub has opinion No in $t=1$, $S_F$ is strictly suboptimal for all values of $p$ and $q$. 
		\item[--] Let the initial opinion of the hub be Yes, assume that the hub is chosen exactly once in time-slots 1 to $T$ and $T=M$. Then, among the class of strategies that influence in $bT$ consecutive time-slots, any strategy that influences the hub outperforms any strategy which does not influence the hub. 
	\end{enumerate}
\end{propn}

Note that in the setting where $T$ scales slower than $M$, the probability of the hub not being selected goes to 1 as the network grows larger. Therefore, in this case, the trends mentioned in the first part of the proposition hold with high probability for large networks. 

In addition, we observe via simulations that the trend mentioned in the second part of the proposition holds even if the initial opinion of the hub is No. We omit these results from this paper due to space constraints.


The key takeaway is that given the special role played by the hub, the performance of any strategy depends heavily on whether it influences in the time-slot in which the hub is chosen or not. As a result, the nature of the optimal strategy for the hub and spoke graph is not the same as for the $d$--regular, Barab\'{a}si Albert, Erd\"{o}s R\'{e}nyi and complete graphs. Unlike Setting \ref{setting:random}, we thus observe that for Setting \ref{setting:fixed}, while the results obtained in \cite{MTNS} extend to $d$--regular, Barab\'{a}si Albert and Erd\"{o}s R\'{e}nyi graphs, they are not completely robust to graph structure.

\section{Proofs} \label{sec:Proofs}

\subsection{Proof of Theorem 1}
Let $X(t)$ denotes the node chosen at time $t$. Thus, $P(X(t)=i)=\frac{1}{M}$. Suppose that the chosen individual $i$ collects $K$ opinions, where $K$ is a random variable taking values in $\{1, \ldots, \infty \}$ such that $P(K=k)=p_k$. Let $n_y(t)$ denote the number of Yes in the collected opinions. Then, given the event $\{ K = k\}$, $\widetilde{\beta_i}(t) = n_y(t)/k$ with probability  $\binom{k}{n_y(t)}(\beta(t))^{n_y(t)}(1 - \beta(t))^{k - n_y(t)}$. We have:
\begin{align*}
P(\xi(t & +1) = 1 |  \mathcal{F}_t) \\
=& \sum_{i = 1}^{M}P(I_i(t+1) = 1 | I_i(t) = 0, X(t) = i) \\
& \times P(I_i(t) = 0 | X(t) = i)P(X(t) = i)\\
\end{align*}
\begin{align*}
=& \frac{1-\beta(t)}{M} \sum_{i = 1}^{M}P(I_i(t+1) = 1 | I_i(t) = 0, X(t) = i) \\
=&\frac{q(1-\beta(t))}{M} \sum_{i = 1}^{M} \sum_{k=1}^M \sum_{n_y(t) = 0}^{k}\frac{n_y(t)}{k}\binom{k}{n_y(t)} \times\\
&\hspace{1in} (\beta(t))^{n_y(t)}(1 - \beta(t))^{k - n_y(t)} p_k   \\
=& (1-\beta(t)) \beta(t) q.
\end{align*}
Similarly, $P(\xi(t+1)=-1|\mathcal{F}_t) = p \beta(t)(1-\beta(t)$. Thus, $E[\xi(t+1)|\mathcal{F}_t] = (q-p)(1-\beta(t))\beta(t)$.
We can now write a stochastic approximation scheme for $\beta(t)$ and obtain the corresponding O.D.E.:
\begin{equation}
    \dot{\beta}(t) = \frac{(q - p)}{M}\beta(t)(1 - \beta(t)).
\end{equation}
Similarly, the corresponding O.D.E. can be obtained for time-slots in which external influence exists. The rest of the proof involves integrating the O.D.E.s over the time periods with and without influences and then comparing the performance at $T$ for the two strategies $S_F$ and $S_L$. We skip the details and refer the readers to \cite{TCNS} since the O.D.E.s turn out to be similar to those obtained for a similar model of opinion evolution on the complete graph in \cite{TCNS}.

\subsection{Proof of Theorem 2}
For a $d$--regular graph, $\tilde{\beta_i}(t) = \beta_i(t) = \dfrac{1}{d} \sum\limits_{j \in N_i} I_j(t)$. Thus,
\begin{equation} \label{BetaBeta_i}
    \beta(t) = \frac{1}{M} \sum_{i \in V} \beta_i(t)  \mbox{ and } \beta_i(t+1) = \beta_i(t) + \frac{1}{d} \sum_{j \in N_i} \xi_j(t+1).
\end{equation}
When there is no external influence we have:
$$ P(\xi_i(t+1) = x | \mathcal{F}_t) = \begin{cases}
p (1 - \beta_i(t)) & \mbox{ for } x = -1 \\
q \beta_i(t) & \mbox{ for } x=1.
\end{cases} $$
From \eqref{BetaBeta_i}, we get:
\begin{eqnarray} \label{SA1}
    \beta(t+1) &=& \beta(t) + \frac{1}{M} \sum_{i \in V}\frac{1}{d}\sum_{j \in N_i}\xi_j(t+1) \nonumber \\
    &=& \beta(t) + \frac{1}{M}\sum_{i \in V} E[\xi_i(t+1)| \mathcal{F}_t]\nonumber \\
    &&+ \frac{1}{M}\sum_{i \in V} ( \xi_i(t+1) - E[\xi_i(t+1)| \mathcal{F}_t]), 
\end{eqnarray}
where $\xi_i(t+1) - E[\xi_i(t+1)| \mathcal{F}_t]$ is a Martingale difference sequence by definition. Thus \eqref{SA1} is a stochastic approximation scheme of the form \eqref{stoch}. The corresponding O.D.E. is given by:
\begin{align*}
    \dot{\beta}(t) &= \frac{1}{M^2}[-\beta(t)p\sum_{i \in V}(1 - \beta_i(t)) + q(1 - \beta(t))q\sum_{i \in V}\beta_i(t)]\\
    &= \frac{(q - p)}{M}\beta(t)(1 - \beta(t)).
\end{align*}
Again, the corresponding O.D.E. for the time-slots with external influence can be written. Then, integrating over the periods with and without influence for strategies $S_F$ and $S_L$ and then comparing the net fraction of the opinion Yes at time $T$ for both the strategies concludes the proof. We refer the readers to \cite{TCNS} for details as the O.D.E.s obtained here are similar.  


\subsection{Proof of Proposition 1}

Let $H$ denote the hub in a hub and spoke graph $G=(V,E)$ with $|V|=M$. Let $I_H(t)$ denote the opinion of the hub in time-slot $t$. For every $i \in V \setminus \{H\}, \beta_i(t) =  \frac{1}{|\mathcal{N}_i|} \sum\limits_{j \in \mathcal{N}_i} I_j(t) = I_H(t)$. Therefore, we have:
\begin{enumerate}[(i)]

	\item If $I_H(t)= I_H(1)=0$, we have $q_t = 0, p_t = p$.
	\item If $I_H(t)= I_H(1)=1$, we have $q_t = q, p_t = 0$.
\end{enumerate}
In either case, when there is external influence, we have $q_t = \tilde{q}, p_t = \tilde{p}$ such that $\tilde{q}>q$ and $\tilde{p}<p$. The proof then follows from Lemma 2 and Theorem 1 \cite{MTNS}.

We now prove the second part of the proposition.
Note that we focus on strategies that influence in $bT$ consecutive time-slots. Let $t_i$ denote the time at which the strategy starts influencing and $t_h$ denote the time at which the hub is chosen. 
We analyze three separate cases by integrating the corresponding O.D.E.s.

\begin{enumerate}
\item[(a)] $0 \leq t_i < t_i + bT < t_h \leq T$: \\
Let $\beta_a(T)$ denote the expected value of $\beta(T)$. 
\begin{eqnarray*}
        \beta_a(T) &=& p\delta(t_h) \beta(t_h) e^{-\frac{p}{M}(T-t_h)} \\
        &&+ (1 - p\delta(t_h))(1 - \delta(t_h) e^{-\frac{q}{M}(T-t_h)}),    
\end{eqnarray*}
where, $\delta(t_h) = \delta(t_0) e^{-\frac{q}{M}(t_h - bT) - \tilde{q}b\frac{T}{M}}$.
\item[(b)] $0 \leq t_i \leq t_h \leq t_i + bT \leq T$ :\\
Let $\beta_b(T)$ denote the expected value of $\beta(T)$. 
\begin{equation*}
    \beta_b(T)= 1 - \delta(t_0) e^{-\frac{T}{M}(q(1-b) + \tilde{q}b)}.
\end{equation*}
\item[(c)] $0 \leq t_h < t_i < t_i + bT \leq T$ :\\
Let $\beta_c(T)$ denote the expected value of $\beta(T)$. 
\begin{eqnarray*}
        \beta_c(T) &=& (1 - p\delta(t_h))(1 - \delta(t_0) e^{-(\frac{T}{M})(q(1-b) + \tilde{q}b)}) \\
        &&+ p\delta(t_h)(1 - (1 - \beta(t_h) e^{(-\frac{p}{M}(t_i - t_h))}e^{-\tilde{q}b\frac{T}{M}}) \\
        && \times e^{-\frac{p}{M}(T-t_i - bT)},
\end{eqnarray*}
where, $  \delta(t_h) = \delta(t_0) e^{-q\frac{t_h}{M}}$.
\end{enumerate}
We now show that $\beta(T)_a \leq \beta(T)_b$. The proof of $\beta(T)_c \leq \beta(T)_b$ is similar. We have:
\begin{align*}
        \beta(T)_a =& p\delta(t_h) [ \beta(t_h) e^{-\frac{p}{M}(T-t_h)} + \delta(t_h) e^{-\frac{q}{M}(T-t_h)} -1 ]  \\
        &+ 1 -  \delta(t_0) e^{-\frac{q}{M}(t_h - bT) - \tilde{q}b\frac{T}{M} -\frac{q}{M}(T-t_h)} \\
        =& \beta(T)_b + p\delta(t_h) [ \beta(t_h) e^{-\frac{p}{M}(T-t_h)} \\
        &+ \delta(t_h) e^{-\frac{q}{M}(T-t_h)} -1 ] . 
\end{align*}
Since $\beta(t_h)+ \delta(t_h)=1$, we get $\beta(T)_a \leq \beta(T)_b$.

\section{Conclusions} \label{sec:Conclusions}

In this work, we focus on characterizing optimal influencing strategies for networks where opinions of individuals evolve based on a limited collection of opinions of individuals the network. We consider two ways in which this collection of opinions is formed.

In the first case, each individual picks a random number of opinions, each chosen uniformly at random to construct the collection of opinions to get influenced by. In this case, we observe that the nature of the optimal influencing strategies is identical to the case where the collection includes all the opinions in the network. In the second case, each individual only takes into account the opinions of her neighbors in a fixed graph. We observe that the nature of the optimal influencing policies is dependent on the nature of the graph. For multiple widely studied graph models, like the $d-$regular, Barab\'{a}si Albert and Erd\"{o}s R\'{e}nyi graphs, the optimal influencing strategies are identical to the case where the collection includes all the opinions in the network.




              \bibliography{ifacconf}

               \end{document}